# Linear magnetoelectric coupling and ferroelectricity induced by the flexomagnetic effect in ferroics

***E.A. Eliseev,[a] M.D. Glinchuk,[a] V. Khist,[a] V.V. Skorokhod,[a] R. Blinc[b] and A.N. Morozovska[*,c]***

[a)]Institute for Problems of Materials Science, National Academy of Sciences of Ukraine, Krjijanovskogo 3, 03142 Kiev, Ukraine

[b)]Jožef Stefan Institute, P. O. Box 3000, 1001 Ljubljana, Slovenia

[c)]V. Lashkarev Institute of Semiconductor Physics, National Academy of Sciences of Ukraine, prospect Nauki 41, 03028 Kiev, Ukraine

## Abstract

Using the symmetry theory we analyze of the flexomagnetic effect in all 90 magnetic classes and showed that 69 of them are flexomagnetic. Then we explore how the symmetry breaking, inevitably present in the vicinity of the surface, changes the local symmetry and thus the form of the flexomagnetic tensors. All possible surface magnetic classes (in the number of 19) were obtained from the 90 bulk magnetic classes for the surface cuts 001, 010 and 100 types. It appeared that all 90 bulk magnetic classes become flexomagnetic, piezomagnetic and piezoelectric in the vicinity of surface.

Using the free energy approach, we show that the flexomagnetic effect leads to a new type of flexo-magnetoelectric (FME) coupling in nanosized and bulk materials, in all spatial regions, where the polarization and (anti)magnetization vectors are spatially inhomogeneous due to external or internal forces.

The linear FME coupling, proportional to the product of the gradients of (anti)magnetization and polarization, flexoelectric and flexomagnetic tensors, is significant in nanosized ferroelectrics-(anti)ferromagnetics, where gradients of the polarization and magnetization obligatory exist. The spontaneous FME coupling induced by the spatial confinement give rise to the size-dependent linear magnetoelectric coupling in nanosized ferroelectrics-(anti)ferromagnetics.

We show that the flexomagnetic effect may lead to improper ferroelectricity in bulk (anti)ferromagnetics via the linear and nonlinear FME coupling. Inhomogeneous spontaneous polarization is induced by the (anti)magnetization gradient, which exists in all spatial regions, where polarization varies and (anti)magnetization vector changes its direction. The gradient can be induced by the surface influence as well as by external strain via e.g. the sample bending.

Generally, the flexomagnetic effect strongly increases the number of the magnetoelectric multiferroic materials of the type-I (ferroelectrics-ferromagnetics with two order parameters - spontaneous polarization and magnetization) and type-II (the materials with inhomogeneous spontaneous magnetization that induces the polarization, or vice versa).

[*] Corresponding author: morozo@i.com.ua

## Introduction

Inhomogeneous strains and electric fields, which can be induced by external forces or exist in the systems with an inhomogeneously distributed bound charge (e.g. varying polarization in the vicinity of the surface), give rise to flexoelectric coupling. Typical example is the flexoelectric effect [1] originating from the coupling of polarization with elastic strain gradient and polarization gradient with elastic strain (direct and converse effects). A detailed theoretical study of the flexoelectric effect was performed by Tagantsev [2, 3, 4], experimental measurements of flexoelectric tensor components in bulk crystals for perovskites were carried out by Ma and Cross [5, 6, 7, 8, 9] and Zubko et al. [10]. Renovation of the theoretical description for the flexoelectric response of different nanostructures starts from Catalan et al [11, 12], Majdoub et al. [13], Kalinin and Meunier [14]. In the papers [10-14] external forces were considered as the sources of the flexoelectric effect. Spontaneous manifestation of flexoelectric effect in ferroelectric nanoparticles due to the intrinsic gradients of order parameter was considered recently by Eliseev et al [15].

Flexomagnetic coupling is much less studied in comparison with the flexoelectric one, only several relevant papers exist [16, 17]. Namely, Lukashev et al. calculated a flexomagnetic coefficient of about 1.95 $\mu_B$ Å for the antiperovskite $Mn_3GaN$ as a coupling between the strain gradient and magnetic dipole moment per Mn atom.

It is worth to underline that the existence of operations of time and/or space inversion alone exclude the flexomagnetic effect. For flexomagnetic effect to exist these operations should be coupled between each other in the material symmetry group. Thus the symmetry consideration for flexomagnetic effect will be similar to the piezomagnetic [18, 19] and magnetoelectric effects [20, 21, 22].

It should be noted, that the inhomogeneous magnetoelectric effect of third order (linear in polarization and quadratic in magnetization, i.e. nonlinear effect) was considered by Bar'yakhtar et al [23]. Recently [24] Pyatakov and Zvezdin applied this approach (which they called flexomagnetoelectric effect) to antiferromagnetics and showed its impact on the magnetoelectric coupling in various multiferroics.

On the other hand the existence of a deformation gradient (direct flexoelectric or flexomagnetic effect) or a magnetization gradient (converse flexomagnetic effect) leads to the absence of space inversion symmetry. The large strain gradient can be experimentally achieved by a sample bending [5-10]. The bending will induce specific magnetic order with noncollinear magnetic structure, which can lead to the appearance of electric polarization. The ferroelectric polarization induced by specific magnetic order was observed for the first time by Kimura et al in $TbMnO_3$ [25]. Since then several type-II multiferroics [26] have been discovered in which magnetic order causes ferroelectric order.

Therefore we are faced with improper ferroelectricity. This type of ferroelectricity was recently discovered in various systems like $RMnO_3$ (R=rare-earth ion), $CuFeO_2$, $CoCr_2O_4$, $Ni_3V_2O_8$, CuO, etc. [27, 28, 29, 30].

In the paper we propose a new mechanism of the ferroelectrics-(anti)ferromagnetics creation with the help of the flexomagnetic effect. Since ferroelectric-(anti)ferromagnetic multiferroics are exceedingly rare (especially in the form of bulk materials) the proposed mechanism besides the fundamental interest can give rise to new technologies and thus could be very important for applications.

The paper is organized in the following way. In the **Section 1** we analyze the possible existence of the flexomagnetic tensor in all 90 bulk magnetic classes and show that 69 of them are flexomagnetic. Then we explore how the spontaneous symmetry breaking, present in the vicinity of surface, changes the local symmetry and thus the form of the flexomagnetic tensors.

In the **Section 2** we show that the flexomagnetic effect leads to a new type of **size-dependent** spontaneous flexo-magnetoelectric (**FME**) coupling in nanosized ferroelectrics-ferromagnetics, where the polarization and magnetization vectors are spatially inhomogeneous. We show that a linear magnetoelectric effect is induced by the linear FME effect in the vicinity of the material surface, as well as the entire nanosystem, once at least one of its sizes is less than 10 nm. In the end of the section we list all the symmetry groups of ferromagnetics-ferroelectrics (multiferroic materials of the type-I), which have a nonzero flexomagnetic effect.

In the **Section 3** we demonstrate that improper ferroelectricity can be induced by the flexomagnetic coupling both in bulk and nanosized ferromagnetics. The main difference is that flexoeffects (and so improper polarization) are spontaneous for the nanosized case, while for a bulk ferromagnetic a pronounced magnetization vector gradient can be induced by external stresses as well as appear in the vicinity of domain walls. At the end of the section we list all the symmetry groups of ferromagnetics-improper ferroelectrics (new multiferroic materials of the type-II), which have a nonzero flexomagnetic effect.

In the **Section 4** we show that the flexomagnetic effect leads to a new type of **size-dependent** spontaneous FME coupling in nanosized ferroelectrics-antiferromagnetics. Similarly to the case of ferroelectrics-ferromagnetics a linear magnetoelectric effect is induced by the linear FME effect in the vicinity of the material surface, as well as the entire nanosystem. The FME coupling strongly affects the order parameter spatial distributions. In the end of the section we list all the symmetry groups of ferroelectrics-antiferromagnetics (multiferroic materials of the type-I) with nonzero flexomagnetic effect.

In the **Section 5** we predict that improper ferroelectricity can be induced by the flexomagnetic coupling both in bulk and nanosized antiferromagnetics. At the end of the section we list all the symmetry groups of antiferromagnetics-improper ferroelectrics (new multiferroic materials of the type-II) with nonzero flexomagnetic effect. We show that the spontaneous FME coupling induced by the spatial confinement give rise to size-dependent linear magnetoelectricity in the type II multiferroic nanosystems.

In the **Section 6** we study the influence of the flexo-magnetoelectric effect on the dielectric and magnetoelectric susceptibility of ferroelectrics-antiferromagnetics and show that the effect of FME coupling between the polarization and magnetization on the tunability and dielectric susceptibility is very high. **Section 7** is discussion and summary.

Let us underline that FME coupling can reflect both the system's response to the external impact and to the intrinsic gradients of polarization and magnetization, existing in nanoferroics due to the surface influence. In the paper we consider both this possibilities. We will show how the flexomagnetic effect strongly increases the number of the magnetoelectric multiferroic materials of the type-I (i.e. ferroelectrics-ferromagnetics with two order parameters - spontaneous polarization and magnetization) and type-II (i.e. the materials with inhomogeneous spontaneous magnetization that induces the polarization, or vice versa).

## 1. Symmetry analyses of the flexomagnetic tensor

Using the symmetry theory we analyze the nonzero components of flexomagnetic tensors for all 90 magnetic classes (see **Supplementary Materials**). We calculate that the flexomagnetic tensor has nonzero components for 69 **bulk** classes, since it is zero for the 21 class with net spatial inversion (including $\bar{1}$, 2/*m*, 2'/*m*', *mmm*, *m*'*m*'*m*, 4/*m*, 4'/*m*, 4/*mmm*, 4/*mm*'*m*', 4'/*mmm*', $\bar{3}$, $\bar{3}m$, $\bar{3}m'$, $6/m$, $6'/m'$, $6/mmm$, $6/mm'm'$, $6'/m'mm'$, *m*3, *m*3*m*, *m*3*m*'). For instance, the flexomagnetic tensor for some cubic symmetries has the form:

(a) 23, m′3: 5 nontrivial elements

$$Q^{(m)}_{ijkl} = \begin{pmatrix} \begin{pmatrix} Q_{1111} & 0 & 0 \\ 0 & Q_{1122} & 0 \\ 0 & 0 & Q_{1133} \end{pmatrix} & \begin{pmatrix} 0 & Q_{1212} & 0 \\ Q_{1221} & 0 & 0 \\ 0 & 0 & 0 \end{pmatrix} & \begin{pmatrix} 0 & 0 & Q_{1221} \\ 0 & 0 & 0 \\ Q_{1212} & 0 & 0 \end{pmatrix} \\ \begin{pmatrix} 0 & Q_{1212} & 0 \\ Q_{1221} & 0 & 0 \\ 0 & 0 & 0 \end{pmatrix} & \begin{pmatrix} Q_{1133} & 0 & 0 \\ 0 & Q_{1111} & 0 \\ 0 & 0 & Q_{1122} \end{pmatrix} & \begin{pmatrix} 0 & 0 & 0 \\ 0 & 0 & Q_{1212} \\ 0 & Q_{1221} & 0 \end{pmatrix} \\ \begin{pmatrix} 0 & 0 & Q_{1221} \\ 0 & 0 & 0 \\ Q_{1212} & 0 & 0 \end{pmatrix} & \begin{pmatrix} 0 & 0 & 0 \\ 0 & 0 & Q_{1212} \\ 0 & Q_{1221} & 0 \end{pmatrix} & \begin{pmatrix} Q_{1122} & 0 & 0 \\ 0 & Q_{1133} & 0 \\ 0 & 0 & Q_{1111} \end{pmatrix} \end{pmatrix} \quad (1.1)$$

(b) $\bar{4}'3m'$, 432, m′3m′: 3 nontrivial elements

$$\begin{pmatrix} \begin{pmatrix} Q_{1111} & 0 & 0 \\ 0 & Q_{1122} & 0 \\ 0 & 0 & Q_{1122} \end{pmatrix} & \begin{pmatrix} 0 & Q_{1212} & 0 \\ Q_{1212} & 0 & 0 \\ 0 & 0 & 0 \end{pmatrix} & \begin{pmatrix} 0 & 0 & Q_{1212} \\ 0 & 0 & 0 \\ Q_{1212} & 0 & 0 \end{pmatrix} \\ \begin{pmatrix} 0 & Q_{1212} & 0 \\ Q_{1212} & 0 & 0 \\ 0 & 0 & 0 \end{pmatrix} & \begin{pmatrix} Q_{1122} & 0 & 0 \\ 0 & Q_{1111} & 0 \\ 0 & 0 & Q_{1122} \end{pmatrix} & \begin{pmatrix} 0 & 0 & 0 \\ 0 & 0 & Q_{1212} \\ 0 & Q_{1212} & 0 \end{pmatrix} \\ \begin{pmatrix} 0 & 0 & Q_{1212} \\ 0 & 0 & 0 \\ Q_{1212} & 0 & 0 \end{pmatrix} & \begin{pmatrix} 0 & 0 & 0 \\ 0 & 0 & Q_{1212} \\ 0 & Q_{1212} & 0 \end{pmatrix} & \begin{pmatrix} Q_{1122} & 0 & 0 \\ 0 & Q_{1122} & \\ 0 & 0 & Q_{1111} \end{pmatrix} \end{pmatrix} \qquad (1.2)$$

(b) $\bar{4}3m$, 4′32′, m′3m: 2 nontrivial elements

$$Q^{(m)}_{ijkl} = \begin{pmatrix} \begin{pmatrix} 0 & 0 & 0 \\ 0 & Q_{1122} & 0 \\ 0 & 0 & -Q_{1122} \end{pmatrix} & \begin{pmatrix} 0 & Q_{1212} & 0 \\ -Q_{1212} & 0 & 0 \\ 0 & 0 & 0 \end{pmatrix} & \begin{pmatrix} 0 & 0 & -Q_{1212} \\ 0 & 0 & 0 \\ Q_{1212} & 0 & 0 \end{pmatrix} \\ \begin{pmatrix} 0 & Q_{1212} & 0 \\ -Q_{1212} & 0 & 0 \\ 0 & 0 & 0 \end{pmatrix} & \begin{pmatrix} -Q_{1122} & 0 & 0 \\ 0 & 0 & 0 \\ 0 & 0 & Q_{1122} \end{pmatrix} & \begin{pmatrix} 0 & 0 & 0 \\ 0 & 0 & Q_{1212} \\ 0 & -Q_{1212} & 0 \end{pmatrix} \\ \begin{pmatrix} 0 & 0 & -Q_{1212} \\ 0 & 0 & 0 \\ Q_{1212} & 0 & 0 \end{pmatrix} & \begin{pmatrix} 0 & 0 & 0 \\ 0 & 0 & Q_{1212} \\ 0 & -Q_{1212} & 0 \end{pmatrix} & \begin{pmatrix} Q_{1122} & 0 & 0 \\ 0 & -Q_{1122} & 0 \\ 0 & 0 & 0 \end{pmatrix} \end{pmatrix} \qquad (1.3)$$

(c) *m*3, *m*3*m*, *m*3*m*′ No flexomagnetic effect

In the **Appendix A** of **Suppl. Materials** we also explore how the symmetry breaking, inevitably present in the vicinity of surface, changes the local symmetry and thus the form of the flexomagnetic tensors. It appears that the 21 bulk class without flexomagnetic effect become flexomagnetic in the vicinity of surface cuts 001, 010 or 100. All possible surface magnetic classes were obtained from the 90 bulk magnetic classes for the cuts 001, 010 or 100. In that way (see **Table A.1** in **Suppl. Materials**) we obtain the following **19 surface magnetic classes**:

***6mm*, 6*m'm'*, 6*'mm'*, 6, 6', 3*m*, 3*m'*, 3, 4*mm*, 4*m'm'*, 4*'mm'*, 4, 4*'*, *mm*2, 2, 2*'*, *m*, *m'*, 1**. (1.4)

As anticipated all classes (1.4) have no inversion center. Higher cuts do not add new surface classes to the 19. So, each of 90 bulk magnetic classes transforms into the one of the 19 surface magnetic classes from Eq.(1.4), all of which are also flexomagnetic, piezomagnetic and piezoelectric due to the absence of the space inversion near the surface [15] and have linear magnetoelectric coupling (except 6′ and 6*'mm'*). So, it is very important that all 90 magnetic classes become flexomagnetic, piezomagnetic, piezoelectric and have linear magnetoelectric coupling (except 6′ and 6*'mm'*) on the nanoscale [18].

## 2. Spontaneous flexo-magnetoelectric coupling in ferromagnetic-ferroelectric nanosystems

For the description of flexo-coupling in the spatially confined **ferromagnetic-ferroelectrics** we will use the Landau-Ginsburg-Devonshire (LGD) phenomenological approach [31, 32, 33, 34, 35, 36, 37] with respect to the surface energy, gradient energy, depolarization or demagnetization fields [15, 18], mechanical stress, flexoelectric and flexomagnetic effects. The total free energy, including the surface contribution, has the form

$$F_V = \int_V \left(g_{FE} + g_{FM} + g_{elast} + g_{striction} + g_{flexo} + g_{ME}\right) d^3r, \tag{2.1a}$$

$$F_S = \int_S d^2r \left( \frac{a_i^S}{2} P_i^2 + K_S (\mathbf{Mn})^2 \right). \tag{2.1b}$$

Here $\mathbf{P}$ is the polarization and $\mathbf{M}$ is the magnetization vector pointing along the magnetic moment direction, $\mathbf{n}$ is the surface normal. The constant $K_S$ in the surface energy is responsible for surface magnetic anisotropy (see e.g. Ref.[38]). Coefficients $a_i^S$ are supposed to be temperature independent. The Gibbs energy density dependence on the order parameters $\mathbf{P}$ and $\mathbf{M}$ is listed below.

The ferroelectric subsystem's contribution is

$$g_{FE} = \frac{a_{ij}^{(e)}(T)}{2} P_i P_j + \frac{a_{ijkl}^{(e)}}{4} P_i P_j P_k P_l + \ldots + \frac{g_{ijkl}^{(e)}}{2} \frac{\partial P_i}{\partial x_j} \frac{\partial P_k}{\partial x_l} - P_i E_i \tag{2.2}$$

Here $\mathbf{E}$ is the electric field. The tensor $g_{ijkl}^{(e)}$ is responsible for polarization gradient energy and should be positively defined.

The contribution of the ferromagnetic subsystem is:

$$g_{FM} = \left( \frac{a_{ij}^{(m)}(T)}{2} M_i M_j + K (\mathbf{Ma})^2 + \frac{g_{ijkl}^{(m)}}{2} \frac{\partial M_i}{\partial x_j} \frac{\partial M_k}{\partial x_l} - \mathbf{HM} \right). \tag{2.3}$$

$K$ is the constant of uniaxial anisotropy, $\mathbf{a}$ is the unit vector pointing along the anisotropy axis, $\mathbf{H}$ is the vector of magnetic field (if any is applied). The tensor $g_{ijkl}^{(m)}$ is responsible for the magnetization gradient energy and should be positively defined. Note, that the coefficient $g_{ijkl}^{(m)}$ is sometimes called "inhomogeneous exchange coupling" [39].

The elastic contribution to the free energy is

$$g_{elast} = \frac{c_{ijkl}}{2} u_{ij} u_{kl} \tag{2.4}$$

Here $u_{ij}$ are strain tensor components, $c_{ijkl}$ are components of elastic stiffness tensor.

The piezoelectric, piezomagnetic, electro- and magneto- striction coupling contribution:

$$g_{striction} = -d_{ijk}^{(e)} P_i u_{jk} - d_{ijk}^{(m)} M_i u_{jk} - q_{ijkl}^{(e)} u_{ij} P_k P_l - q_{ijkl}^{(m)} u_{ij} M_k M_l \qquad (2.5)$$

Here $d_{ijk}^{(e)}$ and $d_{ijk}^{(m)}$ are coupling tensors of piezoelectric and piezomagnetic effects respectively; $q_{ijkl}^{(e)}$ and $q_{ijkl}^{(m)}$ the bulk electro- and magnetostriction coefficients.

The flexomagnetic and flexoelectric coupling energy is:

$$g_{flexo} = \frac{Q_{ijkl}^{(m)}}{2}\left(\frac{\partial u_{ij}}{\partial x_k} M_l - u_{ij}\frac{\partial M_l}{\partial x_k}\right) + \frac{Q_{ijkl}^{(e)}}{2}\left(\frac{\partial u_{ij}}{\partial x_k} P_l - u_{ij}\frac{\partial P_l}{\partial x_k}\right) \qquad (2.6)$$

Tensors $Q_{ijkl}^{(m)}$ and $Q_{ijkl}^{(e)}$ are the forth rank tensors of flexomagnetic and flexoelectric couplings coefficient respectively. Flexoelectric effect exists for arbitrary symmetry.

The magnetoelectric coupling contribution to the free energy (1) has the form

$$g_{ME} = f_{ij} M_i P_j + w_{ijk} M_i P_j P_k + \ldots \qquad (2.7)$$

We included the bilinear coupling term $f_{ij} M_i P_j$ (which exist for 58 magnetic classes in bulk and for ***almost all*** surface magnetic classes inherent to nanosystems). The quadratic terms $\sim M_i M_j P_k$ and $\sim M_i M_j P_k P_l$ are typically small in comparison with the terms linear in magnetization, which are included into Eq.(2.7).

In order to study the flexoelectric and flexomagnetic effects impact on the magnetoelectricity, we neglect depolarization and demagnetization fields (e.g. consider prolate particles with magnetization and polarization directed along the long axes), and surface stress tensor contribution to the free energy expansion (1) as they were studied in details previously [15, 18].

The variation of the bulk (2.1a) and surface free (2.1b) energy on **P**, **M** and $u_{ij}$ gives the equations of state $\delta F_V / \delta M_i = 0$, $\delta F_V / \delta P_i = 0$ and $\delta F_V / \delta u_{ij} = \sigma_{ij}$ in the form of Euler-Lagrange equations ($\sigma_{jk}$ is the stress tensor, δ is the symbol of variational derivative) along with the boundary conditions (see e.g. Refs. [18, 35, 15]). This system of differential equation should be solved along with the equations of mechanical equilibrium $\partial\sigma_{ij}(\mathbf{x})/\partial x_i = 0$ and compatibility equations equivalent to the mechanical displacement vector $u_i$ continuity [40].

The way of how to obtain the analytical solution for mechanical displacements in tubes, wires and pills was derived in Appendix B of Ref. [15]. Using the approach, one can derive the strain tensor components. In the general case the strains contain the terms proportional to the product of flexoelectric or flexomagnetic coefficients on polarization or magnetization gradients respectively as well as the piezoeffect and striction coefficients, and second powers of polarization and magnetization:

$$u_{ij}=-s_{ijmn}\begin{pmatrix} Q^{(e)}_{mnkl}\dfrac{\partial P_k}{\partial x_l}+Q^{(m)}_{mnkl}\dfrac{\partial M_k}{\partial x_l}+d^{(e)}_{mnk}P_i+d^{(m)}_{mnk}M_i \\ +q^{(e)}_{mnkl}\left(P_kP_l-\overline{P_kP_l}\right)+q^{(m)}_{mnkl}\left(M_kM_l-\overline{M_kM_l}\right)\end{pmatrix}. \qquad (2.8)$$

Here $s_{ijkl}$ are components of the elastic compliances tensor. Hereinafter the bar over the letter designates the spatial averaging. Without flexo- and averaged terms the strain (2.8) is the well-known spontaneous strain. The origin of the differences $\left(P_kP_l-\overline{P_kP_l}\right)$ are derived in details by Zhirnov [[41]] and Cao & Cross [42].

In order to demonstrate the pronounced impact of the strains (2.8), let us consider the case of the stress free system, i.e. when the boundary conditions are $\sigma_{ij}n_i\big|_S=0$. Supposing also that in nanosystems with characteristic size about 10 nm the stress is similar to that at the surface, i.e. may be regarded zero everywhere, one could easily find the strain in the explicit form. Substituting the solution for the strain tensor (2.8) into the free energy (2.1) and making Legendre transformations we come to the renormalization of Eq.(2.7) and appearance of a ***new terms*** in the ***magnetoelectric (ME) energy*** and ***flexo-magnetoelectric (FME) energy***:

$$g_{ME}=\left(f_{ij}+s_{wvqs}d^{(e)}_{jwv}d^{(m)}_{isq}\right)M_iP_j+\left(w_{ijk}+s_{wvqs}d^{(m)}_{iwv}q^{(e)}_{sqjk}\right)M_iP_jP_k\ , \qquad (2.9)$$

$$\begin{aligned} g_{FME}=\ & s_{ijqs}Q^{(m)}_{ijkl}Q^{(e)}_{qsnp}\frac{\partial M_k}{\partial x_l}\frac{\partial P_n}{\partial x_p}+s_{ijqs}Q^{(m)}_{ijkl}d^{(e)}_{nsq}P_n\frac{\partial M_l}{\partial x_k}+s_{ijqs}Q^{(e)}_{ijkl}d^{(m)}_{nsq}M_n\frac{\partial P_l}{\partial x_k}+ \\ & +s_{ijqs}Q^{(m)}_{ijkl}q^{(e)}_{qsnp}P_nP_p\frac{\partial M_k}{\partial x_l}+s_{ijqs}Q^{(m)}_{ijkl}q^{(e)}_{qsnp}P_n\frac{\partial P_p}{\partial x_l}M_k+ \\ & +s_{ijqs}q^{(m)}_{ijkl}Q^{(e)}_{qsnp}\frac{\partial P_n}{\partial x_p}M_kM_l+s_{ijqs}Q^{(e)}_{ijkl}q^{(m)}_{qsnp}P_nM_k\frac{\partial M_l}{\partial x_p} \end{aligned} \qquad (2.10)$$

Note, that all ***flexo-magnetoelectric*** terms in Eq.(2.10) ***are absent*** in the initial free energy (2.1). The most important is the ***spontaneous linear flexo-magnetoelectric term***, $g^S_{FME}=s_{ijqs}Q^{(m)}_{ijkl}Q^{(e)}_{qsnp}\dfrac{\partial M_k}{\partial x_l}\dfrac{\partial P_n}{\partial x_p}+s_{ijqs}Q^{(m)}_{ijkl}d^{(e)}_{nsq}P_n\dfrac{\partial M_l}{\partial x_k}+s_{ijqs}Q^{(e)}_{ijkl}d^{(m)}_{nsq}M_n\dfrac{\partial P_l}{\partial x_k}$, that exists under the absence of external factors: magnetic, electric and elastic fields due to the existence of spontaneous magnetization and polarization gradients. Its strength is proportional to the convolution of flexoelectric ($Q^{(e)}_{ijkl}$) and flexomagnetic ($Q^{(m)}_{ijkl}$) tensors, the values of which can be determined experimentally [5-10] or/and calculated from the first principles [16-17]. The term $g^S_{FME}$, as well as the next 2 terms linear in the magnetization $\sim Q^{(m)}_{ijkl} P_n P_p \frac{\partial M_k}{\partial x_l}$ and $\sim Q^{(m)}_{ijkl} P_n \frac{\partial P_p}{\partial x_l} M_k$, exist in the materials with nonzero flexomagnetic tensor $Q^{(m)}_{ijkl}$. These terms are responsible for appearance of inhomogeneous polarization and magnetization in spatially modulated ferromagnetics. The nonlinear magnetization terms $\sim P_n M_k \frac{\partial M_l}{\partial x_p}$ and $\sim \frac{\partial P_n}{\partial x_p} M_k M_l$ exist in the materials of arbitrary symmetry, since the flexoelectric tensor $Q^{(e)}_{ijkl}$ and magnetostriction tensor $q^{(m)}_{ijnp}$ have nonzero components for arbitrary symmetry.

In the **Table 1** we listed the symmetry groups of ferromagnetics-ferroelectrics, which have nonzero flexomagnetic effect (some of $Q^{(m)}_{ijkl} \neq 0$). Note, that all 13 ferromagnetic-ferroelectric groups from the table can be the surface ones as included in the list (1.4).

All ferromagnetic-ferroelectric groups, listed in the **Table 1**, are linear magnetoelectrics, piezomagnetics and piezoelectrics ($d^{(e)}_{ijk} \neq 0$ and $d^{(m)}_{ijk} \neq 0$) both in the bulk and near the surface.

Note, that the number of the nonzero tensor components is always higher (up to the several times) than the number of the nontrivial components. The full form of the flexomagnetic tensors are listed in the **Suppl. Materials**.

**Table 1.** Ferromagnetics-ferroelectrics with flexomagnetic effect, which are multiferroics of the type I

| **Point symmetry group** | **Magnetic symmetry group** | **Number of the tensors nontrivial components *** | | | |
|---|---|---|---|---|---|
| | | flexo-magnetic | linear magneto-electric | piezo-magnetic | piezo-electric |
| 1 | 1 | 54 | 9 | 18 | 18 |
| 2 | 2 | 28 | 5 | 8 | 8 |
| | 2' | 26 | 4 | 10 | |
| *m* | *m* | 26 | 4 | 8 | 10 |
| | *m'* | 28 | 5 | 10 | |
| *mm*2 | *m'm'*2 | 15 | 3 | 5 | 5 |
| | *mm'*2' | 13 | 2 | 3 | |
| 4 | 4 | 14 | 2 | 4 | 4 |
| 4*mm* | 4*m'm'* | 8 | 2 | 3 | 3 |
| 3 | 3 | 18 | 2 | 6 | 6 |
| 3*m* | 3*m'* | 11 | 1 | 2 | 4 |
| 6 | 6 | 12 | 2 | 4 | 4 |
| 6*mm* | 6*m'm'* | 7 | 2 | 3 | 3 |

Similarly to the flexoelectric effect considered in Ref. [15], the flexomagnetic coupling and magnetostriction effects lead to the renormalization of the gradient terms in Eq.(2.2) and (2.3) as

$$g^{(Rm)}_{klpn} = g^{(m)}_{klpn} - Q^{(m)}_{ijkl} s_{ijsq} Q^{(m)}_{sqpn}, \quad g^{(Re)}_{klpn} = g^{(e)}_{klpn} - Q^{(e)}_{ijkl} s_{ijsq} Q^{(e)}_{sqpn}. \qquad (2.11a)$$

Piezoelectric and piezomagnetic coupling leads to the renormalization of the expansion series coefficients in Eq.(2.2) and (2.3):

$$\tilde{a}_{ij}^{(e)} = a_{ij}^{(e)} - \frac{1}{2} d_{ilp}^{(e)} s_{lpkm} d_{jkm}^{(e)}, \qquad \tilde{a}_{ij}^{(m)} = a_{ij}^{(m)} - \frac{1}{2} d_{ilp}^{(m)} s_{lpkm} d_{jkm}^{(m)}. \qquad (2.11b)$$

To study the spontaneous linear ***flexo-magnetoelectric*** coupling in ferroelectrics-ferromagnetics, let us consider the model of ***one-dimensional*** distributions of the one component polarization and magnetization inside an ***ultrathin nanotube*** with inner radius $R_i$ and outer radius $R_o$ where the tube thickness $h = R_o - R_i$ is very small in comparison with the average tube radius $R = 0.5(R_o + R_i)$ (see **Fig.1a**).

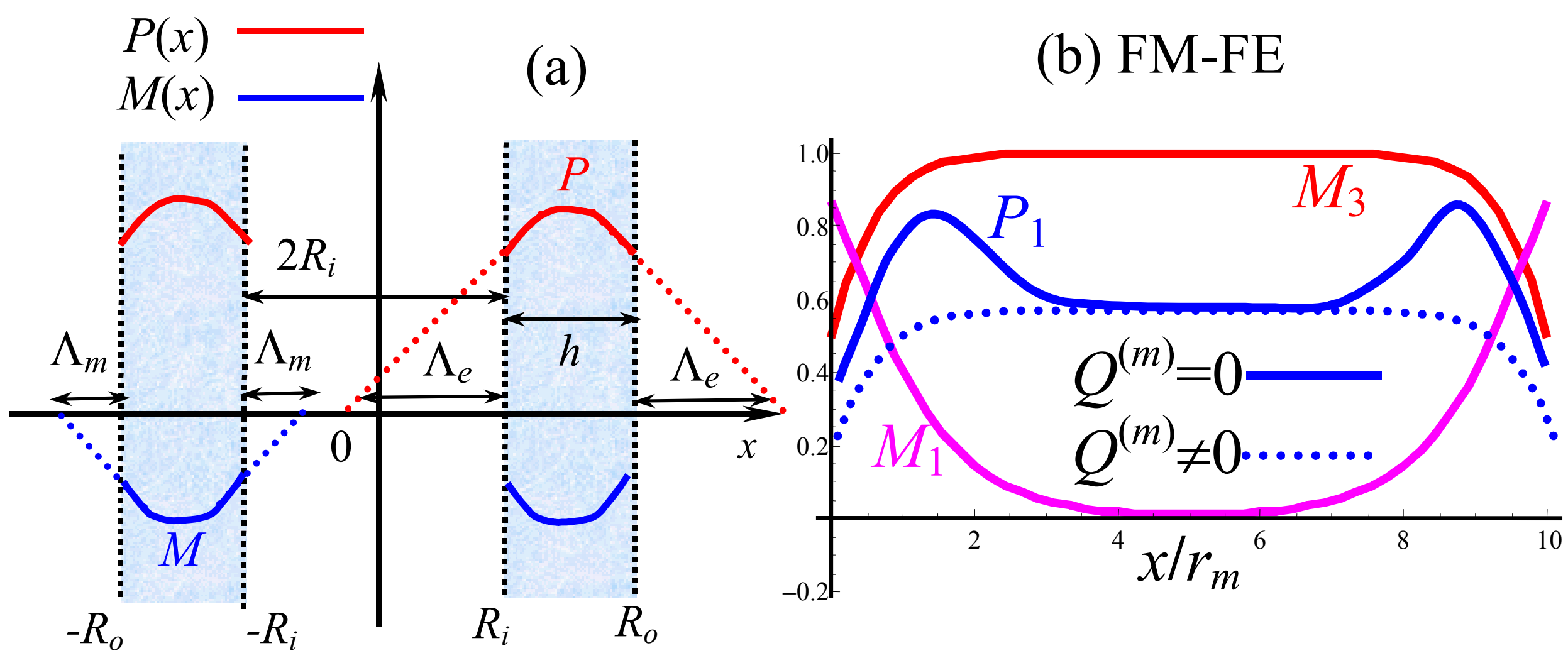

**Fig. 1.** (a) Sketch of one-dimensional polarization and magnetization distributions (solid curves) inside the nanotube: $\Lambda_e$ and $\Lambda_m$ are corresponding extrapolation lengths, which geometrical sense is the distance on x-axes cut by the tangent line to the points $x = \pm R_{i,o}$. (b) Spontaneous magnetization $M_{1,3}(x)$ and polarization $P_1(x)$ normalized distributions in a ferromagnetic-ferroelectric nanotube. Solid curve is $P_1(x)$ calculated with flexomagnetic effect $Q^{(m)}\neq 0$. Dotted curve is $P_1(x)$ calculated without flexomagnetic effect $Q^{(m)}=0$.

This simple model can be useful, because it allows analytical calculations of the average properties, which are measured by most of conventional experimental methods.

For thin tubes ($h << R$) we estimate the average value of the gradients products:

$$\overline{\frac{\partial P}{\partial x}\frac{\partial M}{\partial x}} \approx \frac{1}{h}\int_{R_i}^{R_0} \frac{\partial P(x)}{\partial x}\frac{\partial M(x)}{\partial x} dx \sim \frac{2 r_e r_m \overline{MP}}{(r_e + r_m)(r_e + \Lambda_e)(r_m + \Lambda_m)h}, \qquad (2.12)$$

where the electric and magnetic correlation lengths are introduced as $r_e = \sqrt{\tilde{g}^{(e)} / \left| a^S \right|}$ and $r_m = \sqrt{\tilde{g}^{(m)} / |K|}$, $\Lambda_e = \tilde{g}^{(e)} / a^S$ and $\Lambda_m = \tilde{g}^{(m)} / K_S$ are corresponding extrapolation lengths, which origin and peculiarities related with flexoeffect are considered in e.g. Ref. [15]. For ferroelectrics the extrapolation length values calculated from the first principles are $\Lambda_k^e \sim 1$ nm [43].

Allowing for the fact, that the extrapolation length is proportional to the gradient value given by Eq.(2.11), its renormalization due to the flexoeffects should also be taken into account. Sometimes it may lead to the disappearance of the size-induced ferroelectric/ferromagnetic phase transition [15].

So, the spontaneous FME coupling induced by the spatial confinement give rise to ***additional size-dependent*** linear magnetoelectricity in the ferromagnetic-ferroelectric nanosystems. FME coupling affects the polarization distribution (see **Fig.1b**).

Using the approximate expressions like (2.12), the spontaneous linear FME coupling energy inherent to the nanosystem is:

$$\overline{g_{FME}^{S}} = \frac{1}{V} \int_V g_{FME}^S d^3 r \approx s_u \left( \frac{2 r_e r_m Q^{(m)} Q^{(e)}}{(r_e + r_m)(r_e + \Lambda_e)(r_m + \Lambda_m)} + \frac{r_m Q^{(m)} d^{(e)}}{(r_m + \Lambda_m)} + \frac{r_e Q^{(e)} d^{(m)}}{(r_e + \Lambda_e)} \right) \frac{\overline{MP}}{h}. \quad (2.13)$$

In Eq.(2.13) we omit all subscripts in the compliances, flexoeffect, piezomagnetic and piezoelectric tensors for the sake of simplicity. It is seen from Eq.(2.13) and **Fig.2a** that the coupling term has a pretty strong thickness dependence, namely its strength is inversely proportional to the tube thickness *h*. It is seen that the coupling decreases with magnetization and polarization extrapolation lengths $\Lambda_m$ and $\Lambda_e$, since the gradients become smaller with $\Lambda_{e,m}$ increase. The higher is $\Lambda_{e,m}$ the smaller will be FME coupling.

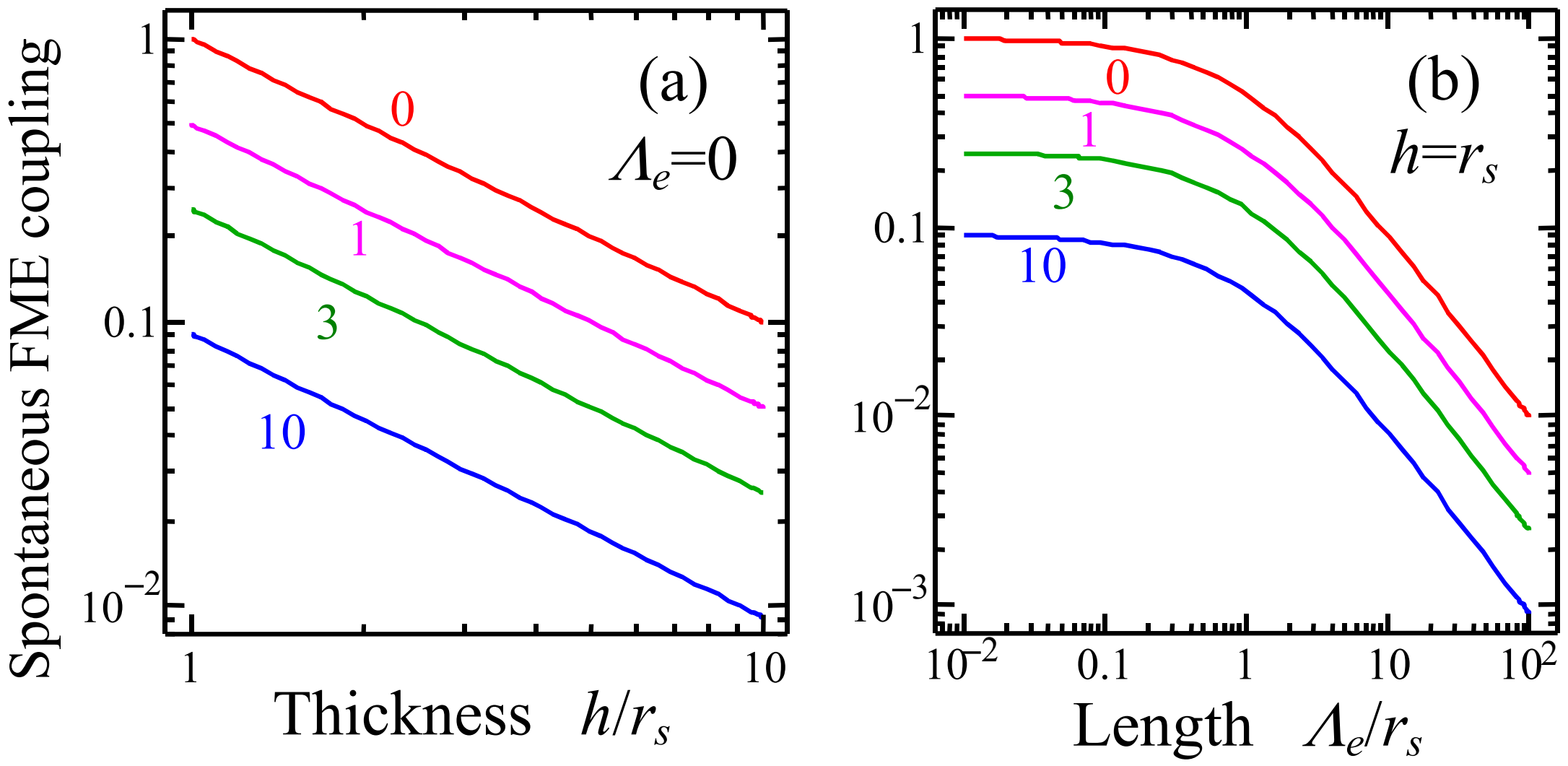

**Fig. 2.** Relative FME coefficients $\frac{g^{S}_{FME}}{Q^{(m)}s_u Q^{(e)}}$ as a function of the nanotube thickness $h$ (a) and magnetic extrapolation length $\Lambda_m$. The characteristic length $r_S = \frac{2r_e r_m}{(r_e + r_m)}$. Different curves in plots (a, b, c, d, f) correspond to different values of $\Lambda_m/r_m$ =0, 1, 3, 10 (marked near the curves). Other parameters: $d^{(e)} = 0$, $d^{(m)} = 0$, $\Lambda_e/r_e$ =0 (a) and $h = r_S$ (b).

It is worth to underline that homogeneous ME and inhomogeneous FME energy described by Eq.(2.9) and Eq.(2.10) respectively could appear also in bulk systems with nonzero magnetoelectric, piezo- and flexoeffects, the latter being induced by external forces.

## 3. Improper ferroelectricity induced by the flexo-magnetoelectric coupling in ferromagnetics

In this section we will show that the application of external stresses to the **bulk** or internal stresses existing in **confined** ferromagnets with flexomagnetic effect due to the FME coupling can induce the electric polarization, i.e. an ***improper ferroelectric phase***. Schematically the improper ferroelectricity origin can be explained in the following way (see **Fig.3a**). Elastic strain via the flexoelectric and piezoelectric effects induces the polarization vector $P_i \sim Q^{(e)}_{ijkl}\frac{\partial u_{kl}}{\partial x_j} + d^{(e)}_{ikl}u_{kl}$. Since the strain can also include, piezoelectric, piezomagnetic and flexomagnetic contributions: $u_{kl} \sim d^{(m)}_{nkl}M_n + Q^{(m)}_{inkl}\frac{\partial M_i}{\partial x_n}$,

The polarization vector becomes

$$P_i \sim Q^{(e)}_{ijkl} d^{(m)}_{nkl} \frac{\partial M_n}{\partial x_j} + Q^{(e)}_{ijkl} Q^{(m)}_{ijkl} \frac{\partial^2 M_i}{\partial x_j \partial x_n} + d^{(e)}_{ikl} Q^{(m)}_{jnkl} \frac{\partial M_j}{\partial x_n} + d^{(e)}_{ikl} d^{(m)}_{nkl} M_n .$$

More rigorously, equation of state for polarization vector follows from the variation of the free energy (2.1) $\delta F_V / \delta P_i = 0$. Variation of the linear FME and ME coupling energy

$$g_{ME} + g_{FME} = \begin{pmatrix} s_{ijqs} Q^{(m)}_{ijkl} Q^{(e)}_{qsnp} \dfrac{\partial M_k}{\partial x_l} \dfrac{\partial P_n}{\partial x_p} + s_{ijqs} Q^{(m)}_{ijkl} d^{(e)}_{nsq} P_n \dfrac{\partial M_l}{\partial x_k} \\ + s_{ijqs} Q^{(e)}_{ijkl} d^{(m)}_{nsq} M_n \dfrac{\partial P_l}{\partial x_k} + \left( f_{ij} + s_{wvqs} d^{(e)}_{jwv} d^{(m)}_{jsq} \right) M_i P_j \end{pmatrix} \qquad (3.1)$$

leads to the built-in electric field appearance in the right-hand side of the equation for polarization:

$$a^{(e)}_{ij}(T) P_j + a^{(e)}_{ijkl} P_j P_k P_l + \ldots = E^{FME}_n + E^{ME}_n , \qquad (3.2a)$$

$$E^{FME}_n = -s_{ijqs} Q^{(m)}_{ijkl} Q^{(e)}_{qsnp} \frac{\partial^2 M_k}{\partial x_l \partial x_p} + s_{ijqs} Q^{(m)}_{ijkl} d^{(e)}_{nsq} \frac{\partial M_l}{\partial x_k} - s_{ijqs} Q^{(e)}_{ijkn} d^{(m)}_{lsq} \frac{\partial M_l}{\partial x_k} . \qquad (3.2b)$$

$$E^{ME}_n = \left( f_{nj} + s_{wvqs} d^{(e)}_{nwv} d^{(m)}_{isq} \right) M_i . \qquad (3.2c)$$

The inhomogeneous built-in field (3.2b) has 3 terms: the first one, $-s_{ijqs} Q^{(m)}_{ijkl} Q^{(e)}_{qsnp} \frac{\partial^2 M_k}{\partial x_l \partial x_p}$, exists in ferromagnetics with nonzero flexomagnetic effect, the second one, $s_{ijqs} Q^{(m)}_{ijkl} d^{(e)}_{nsq} \frac{\partial M_l}{\partial x_k}$, exists in ferromagnetics with nonzero flexomagnetic and piezoelectric effect (but the latter is nonzero in the vicinity of any surface due to the absence of the space inversion here [15]), the third one, $-s_{ijqs} Q^{(e)}_{ijkn} d^{(m)}_{lsq} \frac{\partial M_l}{\partial x_k}$, exists in ferromagnetics with nonzero flexoelectric and piezomagnetic effect, while in Ref.[15] we show that almost all nanomaterials are piezomagnetic in the vicinity of surface. The built-in field $E^{ME}_n$ given by Eq.(3.2c) originates from the renormalized linear ME coupling, that is very rare for bulk materials. The built-in fields induce the polarization under the absence of external electric field (see **Fig.3b**).

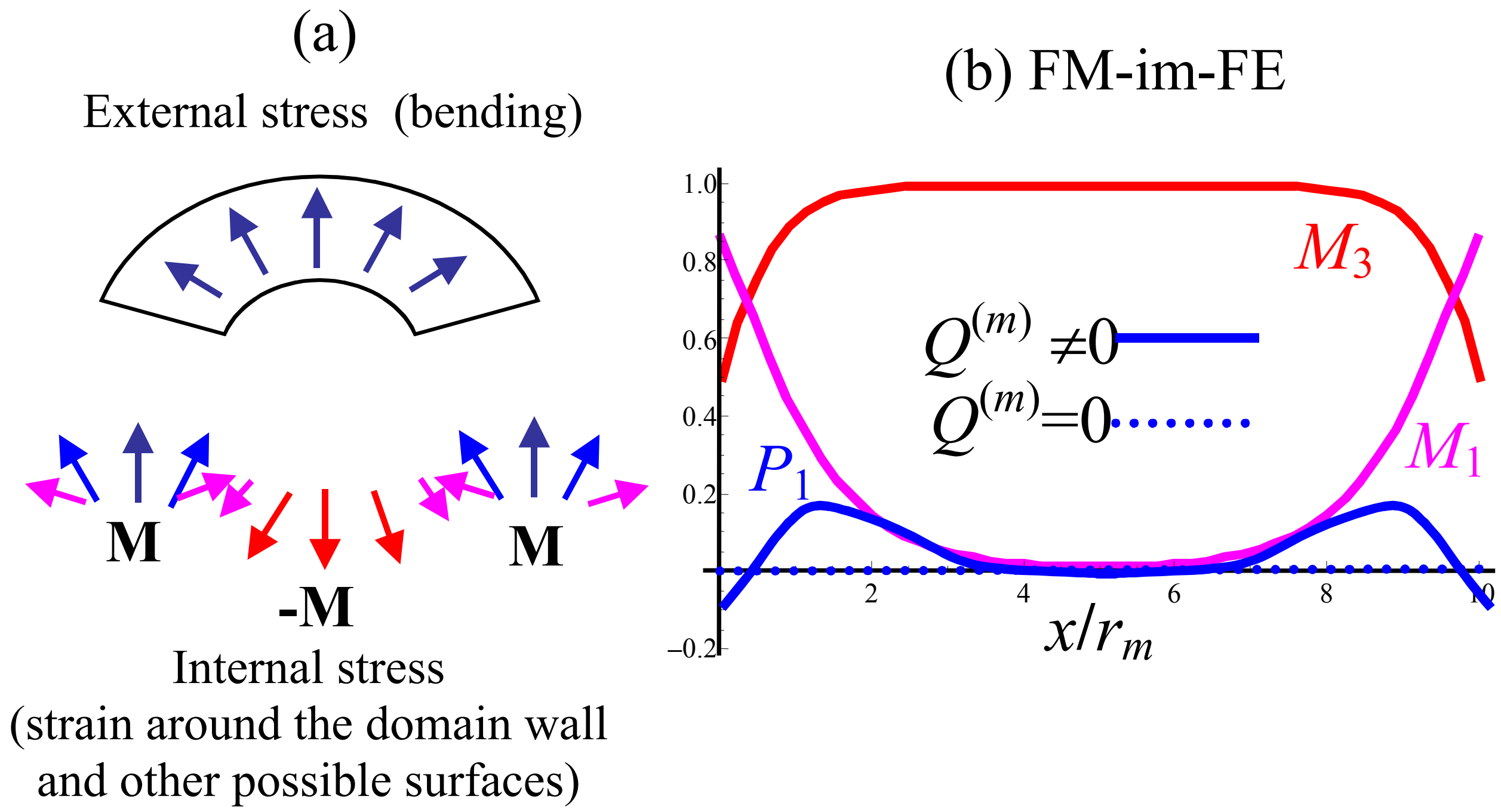

**Fig. 3.** (a) Sketch of the inhomogeneous external stress (bending) and internal stress (e.g. the strains around domain walls or other possible surfaces) originating from flexomagnetic and piezomagnetic effects in ferromagnetics. (b) Inhomogeneous polarization distribution $P_1(x)$ induced in the type II multiferroic by the electric built-in field originating from FME coupling. Inhomogeneous magnetization components $M_{1,3}(x)$ induce the polarization $P_1(x)$ in ferromagnetics. Solid curve is $P_1(x)$ calculated with flexomagnetic effect $Q^{(m)}\neq 0$. Dotted line is zero polarization without flexomagnetic effect $Q^{(m)}=0$.

Therefore the appearance of improper ferroelectricity and linear ME coupling induced by the flexoeffects should exist both in bulk and nanosized ferromagnetics. The main difference is that flexoeffects (and so improper polarization) are spontaneous for the nanosized case, because the magnetization vector rotation (gradient) appears without external forces near the surface. However, for a bulk ferromagnetic material, a pronounced magnetization vector gradient can be induced by external stresses. It may as well appearin the vicinity of Bloch-type domain walls.

To be complete, in the **Table 2** we listed all possible symmetry groups of ferromagnetics, which can be improper ferroelectrics due to nonzero flexomagnetic effect (some of $Q^{(m)}_{ijkl} \neq 0$). Note, that two of the groups have no linear magnetoelectric effect (despite being piezomagnetics and piezoelectrics). So, the induced FME coupling gives rise to the linear magnetoelectricity in type II multiferroics.

**Table 2.** Ferromagnetics-improper ferroelectrics due to the flexomagnetic effect (multiferroics of the type II)

| *) Point symmetry group | Magnetic symmetry group | Number of the tensors nontrivial components * | | | |
|---|---|---|---|---|---|
| | | flexo-magnetic | linear magneto-electric | piezo-magnetic | piezo-electric |
| 222 | 22'2' | 13 | 2 | 5 | 3 |
| $\bar{4}$ | $\bar{4}$ | 14 | 2 | 4 | 4 |
| $\bar{4}2m$ | $\bar{4}2'm'$ | 7 | 1 | 3 | 2 |
| 422 | $42'2'$ | 6 | 1 | 3 | 1 |
| 32 | 32' | 9 | 1 | 4 | 2 |
| $\bar{6}$ | $\bar{6}$ | **6** | **0** | 4 | 2 |
| $\bar{6}m2$ | $\bar{6}m'2'$ | **3** | **0** | 3 | 1 |
| 622 | 62'2' | 5 | 2 | 3 | 1 |

*) Note, that none of the groups from the table can be the surface one (see (1.4))

Using the proposed approach one can show that improper magnetization can be induced by the flexomagnetic coupling both in bulk and nanosized ferroelectrics.

## 4. Spontaneous flexo-magnetoelectric coupling in nanosized antiferromagnetic-ferroelectrics

In all subsequent sections we will us consider antiferromagnetics. First of all let us underline the difference in the ferromagnetics and antiferromagnetics phenomenological description.

Below we consider the ferroelectric-antiferromagnetic with two sublattices *a* and *b*. Antiferromagnetic order parameter $\mathbf{L}=\left(\mathbf{M}^{(a)}-\mathbf{M}^{(b)}\right)/2$ transforms as an axial vector relatively symmetry operations of the atoms in one sublattice and changes it sign at $a\leftrightarrow b$. It is known [44] that the piezomagnetic tensor sign follows the sign of **L**. In such a case nonzero piezomagnetic tensor components $d_{ijk}^{(m)}$ will have the same indexes for the free energy in the form $d_{ijk}^{(m)}L_i u_{jk}$ or $\tilde{d}_{ijk}^{(m)}H_i u_{jk}$, although their values can be different. The same reasons can be valid for linear magnetoelectric tensors, which can be written in the form $f_{ij}L_iP_j$ or $\tilde{f}_{ij}H_iE_j$, as well as for the flexomagnetic tensors, which can be written in the form $Q_{ijkl}^{(m)}\frac{\partial u_{ij}}{\partial x_k}L_l$ or $\tilde{Q}_{ijkl}^{(m)}\frac{\partial u_{ij}}{\partial x_k}H_l$. Therefore the tensors of all effects linear on **L** should change their sign simultaneously with **L**, i.e. under the interchanging the atoms in the different sublattices.

Since magnetic field **H** and electric field **E** are regarded absent below, the ferromagnetic order parameter $\mathbf{M}=\left(\mathbf{M}^{(a)}+\mathbf{M}^{(b)}\right)/2$ is also absent, and the antiferromagnetic free energy has the form:

$$F_V = \int_V \left( g_{FE} + g_{AFM} + g_{elast} + g_{striction} + g_{flexo} \right) d^3 r, \tag{4.1a}$$

$$F_S = \int_S d^2 r \left( \frac{a_i^S}{2} P_i^2 + \left( 2K_S - \tilde{K}_S \right) (\mathbf{Ln})^2 \right), \tag{4.1b}$$

Here $\mathbf{n}$ is the surface normal, $K_S$ is a sub-lattice surface anisotropy, $\tilde{K}_S$ is the inter-lattices surface anisotropy [45, 46], the ferroelectric part $g_{FE}$ is given by Eq.(2.2) at **E**=0.

The antiferromagnetic contribution is

$$g_{AFM} = -J \cdot \mathbf{L}^2 + \left( 2K - \tilde{K} \right) L_3^2 + \left( g_{ijkl}^{(m)} - \tilde{g}_{ijkl}^{(m)} \right) \frac{\partial L_i}{\partial x_j} \frac{\partial L_k}{\partial x_l}. \tag{4.2}$$

Here $g_{ijkl}^{(m)}$ is a sub-lattice inhomogeneous exchange coefficient, $\tilde{g}_{ijkl}^{(m)}$ is the inter-lattices inhomogeneous exchange coefficient, $K$ is sub-lattice bulk anisotropy, $\tilde{K}_S$ is an inter-lattice bulk anisotropy, $J$ is the sublattices exchange coupling constant. The condition $J>0$ is necessary for antiferromagnetic state, $\mathbf{M}^{(a)} = -\mathbf{M}^{(b)}$, to be stable at zero and small magnetic fields.

The elastic contribution to the free energy $g_{elast}$ is given by Eq.(2.4). Piezoelectric, piezomagnetic, electro- and magneto- striction contribution:

$$g_{striction} = \left( -d_{ijk}^{(e)} P_i u_{jk} - d_{ijk}^{(m)} L_i u_{jk} - q_{ijkl}^{(e)} u_{ij} P_k P_l - \left( 2q_{ijkl}^{(m)} - \tilde{q}_{ijkl}^{(m)} \right) u_{ij} L_k L_l \right) \tag{4.3}$$

$q_{ijkl}^{(m)}$ is a sub-lattice electrostriction, $\tilde{q}_{ijkl}^{(m)}$ is inter-lattices bulk electrostriction.

Flexo-magnetic and flexoelectric coupling:

$$g_{flexo} = \frac{Q_{ijkl}^{(m)}}{2} \left( \frac{\partial u_{ij}}{\partial x_k} L_l - u_{ij} \frac{\partial L_l}{\partial x_k} \right) + \frac{Q_{ijkl}^{(e)}}{2} \left( \frac{\partial u_{ij}}{\partial x_k} P_l - u_{ij} \frac{\partial P_l}{\partial x_k} \right) \tag{4.4}$$

Note, that for the considered case of commutative sublattices we put $Q_{ijkl}^{(a)} \equiv Q_{ijkl}^{(b)} \equiv Q_{ijkl}^{(m)}$.

In contrast to the general free energy (3.1), listed for ferroelectrics-ferromagnetics in the previous section, below we mainly consider the ferroelectrics-antiferromagnetics of definite symmetry. We suppose that the high temperature paraelectrics and paramagnetic phase has the surface symmetry group 4m′m′ (that by the way corresponds to e.g. the m′3m′-bulk symmetry) allowing both the flexomagnetic and bilinear magnetoelectric coupling. Also we consider the case of only one component of ferroelectric polarization, $P_1$, two components of antimagnetization vector $L_{1,3}$ and magnetic anisotropy axis along the axis $z$ (see **Fig. 4a**).

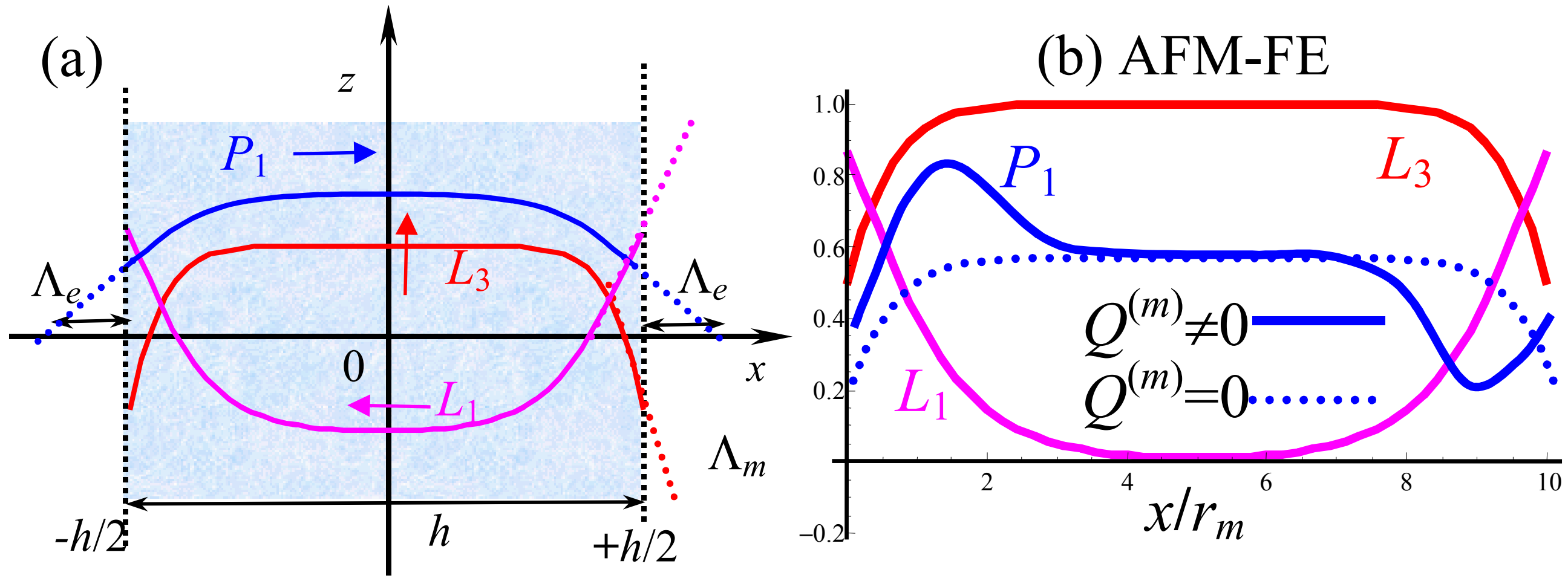

**Fig. 4.** (a) Sketch of one-dimensional antiferromagnetic order parameter **L** and polarization distributions inside the film: $\Lambda_e$ and $\Lambda_m$ are corresponding extrapolation lengths, which geometrically is the distance on x-axes cut by the tangent line to the points $x=\pm h/2$. (b) Inhomogeneous normalized antimagnetization $L_{1,3}(x)$ and polarization $P_1(x)$ distribution in the antiferromagnetic-ferroelectric film. Solid curve is $P_1(x)$ calculated with flexomagnetic effect $Q^{(m)}\neq 0$. Dotted curve is $P_1(x)$ calculated without flexomagnetic effect $Q^{(m)}=0$.

Below we consider an ***ultrathin antiferromagnetic-ferroelectric film on a matched substrate*** (i.e. misfit strain is negligibly small). For this case the solution is $u_{11}=u_{22}=u_{23}=0$ and $\sigma_{13}=0$, $\sigma_{23}=0$, $\sigma_{33}=0$ (see **Appendix B, Suppl. Materials**), while

$$u_{33}=\frac{1}{c_{11}}\left(q_{12}^{(e)}P_1^2+d_{33}^{(m)}L_3+\left(2q_{11}^{(m)}-\tilde{q}_{11}^{(m)}\right)L_3^2+Q_{11}^{(m)}\frac{\partial L_3}{\partial x_3}\right) \tag{4.5a}$$

$$u_{13}=\frac{1}{c_{44}}\left(d_{15}^{(e)}P_1+Q_{44}^{(e)}\frac{\partial P_1}{\partial x_3}+d_{15}^{(m)}L_1+\left(2q_{44}^{(m)}-\tilde{q}_{44}^{(m)}\right)L_1L_3+Q_{44}^{(m)}\frac{\partial L_1}{\partial x_3}\right) \tag{4.5b}$$

The substitution of the elastic solution (4.5) into the free energy (4.1) and consequent Legendre transformation leads to the following expression for the free energy:

$$\tilde{F}=\int_V\left(\tilde{g}_{FE}+\tilde{g}_{AFM}+\tilde{g}_{ME}+\tilde{g}_{FME}\right)d^3r+\int_S d^2r\left(\frac{a_i^S}{2}P_i^2+\left(2K_S-\tilde{K}_S\right)(\mathbf{nL})^2\right) \tag{4.6}$$

The ferroelectric contribution acquires the form:

$$\tilde{g}_{FE}=\frac{1}{2}\left(a_1^{(e)}-\frac{\left(d_{15}^{(e)}\right)^2}{c_{44}}\right)P_1^2+\left(\frac{a_{11}^{(e)}}{4}-\frac{\left(q_{12}^{(e)}\right)^2}{2c_{11}}\right)P_1^4+\ldots+\left(\frac{g_{44}^{(e)}}{2}-\frac{\left(Q_{44}^{(e)}\right)^2}{2c_{44}}\right)\left(\frac{\partial P_1}{\partial x_3}\right)^2 \tag{4.7}$$

For 4m′m′ symmetry the antiferromagnetic contribution acquires the form:

$$\widetilde{g}_{AFM}=\begin{pmatrix} -J\left(L_1^2+L_3^2\right)+\left(2K-\widetilde{K}-\frac{\left(d_{33}^{(m)}\right)^2}{2c_{11}}\right)L_3^2-\frac{\left(2q_{44}^{(m)}-\widetilde{q}_{44}^{(m)}\right)Q_{44}^{(m)}}{c_{44}}L_1L_3\frac{\partial L_1}{\partial x_3}-\frac{d_{33}^{(m)}Q_{11}^{(m)}}{c_{11}}L_3\left(\frac{\partial L_3}{\partial x_3}\right) \\ +\left(g_{44}^{(m)}-\widetilde{g}_{44}^{(m)}-\frac{\left(Q_{44}^{(m)}\right)^2}{2c_{44}}\right)\left(\frac{\partial L_1}{\partial x_3}\right)^2+\left(g_{11}^{(m)}-\widetilde{g}_{11}^{(m)}-\frac{\left(Q_{11}^{(m)}\right)^2}{2c_{11}}\right)\left(\frac{\partial L_3}{\partial x_3}\right)^2-\frac{d_{15}^{(m)}Q_{44}^{(m)}}{c_{44}}L_1\frac{\partial L_1}{\partial x_3} \end{pmatrix} \tag{4.8}$$

New ***ME*** and ***FME coupling*** terms are

$$\widetilde{g}_{ME}=-\frac{d_{15}^{(e)}d_{15}^{(m)}}{c_{44}}P_1L_1-\frac{q_{12}^{(e)}d_{33}^{(m)}}{c_{11}}P_1^2L_3-\frac{\left(2q_{44}^{(m)}-\widetilde{q}_{44}^{(m)}\right)d_{15}^{(e)}}{c_{44}}P_1L_1L_3-\frac{q_{12}^{(e)}\left(2q_{11}^{(m)}-\widetilde{q}_{11}^{(m)}\right)}{c_{11}}P_1^2L_3^2, \tag{4.9a}$$

$$\widetilde{g}_{FME}=\frac{1}{c_{44}}\begin{pmatrix} -Q_{44}^{(e)}Q_{44}^{(m)}\frac{\partial L_1}{\partial x_3}\frac{\partial P_1}{\partial x_3}-Q_{44}^{(e)}d_{15}^{(m)}L_1\frac{\partial P_1}{\partial x_3}-d_{15}^{(e)}Q_{44}^{(m)}\frac{\partial L_1}{\partial x_3}P_1 \\ +Q_{44}^{(e)}\left(\widetilde{q}_{44}^{(m)}-2q_{44}^{(m)}\right)\left(\frac{\partial P_1}{\partial x_3}\right)L_1L_3-\frac{q_{12}^{(e)}Q_{11}^{(m)}}{c_{11}}P_1^2\frac{\partial L_3}{\partial x_3} \end{pmatrix}. \tag{4.9b}$$

Note that the novel linear and nonlinear terms (4.9b) can exist in ferroelectrics-antiferromagnetics with developed gradient of polarization **P** and/or anti-magnetization **L**. The FME coupling affects the order parameter spatial distributions as shown in **Fig.4b**. In particular, pronounced maxima appear on the polarization distribution in the regions, where the gradient of the anti-magnetization **L** exists (e.g. near the film surfaces, where the $L_1$ and $L_3$ change their values due to the rotation of vector **L**).

The simple model with extrapolation length allows analytical calculations of the average properties, which are measured conventional experimental methods. For thin films of thickness $h$ we estimated the average values:

$$\overline{\left(\frac{\partial P_1}{\partial x}\right)\left(\frac{\partial L_1}{\partial x}\right)}=\frac{1}{h}\int_{-h/2}^{-h/2}\frac{\partial P_1(x)}{\partial x}\frac{\partial L_1(x)}{\partial x}dx\approx\frac{2r_er_m\overline{P_1L_1}}{(r_e+r_m)(r_e+\Lambda_e)(r_m+\Lambda_m)h}, \tag{4.10a}$$

$$\overline{\left(\frac{\partial P_1}{\partial x}\right)L_1L_3}=\frac{1}{h}\int_{-h/2}^{-h/2}\frac{\partial P_1(x)}{\partial x}L_1(x)L_3(x)dx\approx\frac{r_e\overline{P_1L_1L_3}}{(r_e+\Lambda_e)h}, \tag{4.10b}$$

where the electric correlation and extrapolation lengths are introduced as $r_e=\sqrt{g^{(e)}/\left|a^S\right|}$ and $\Lambda_e=g^{(e)}/a^S$.

Using the approximate expression (4.10), the spontaneous linear FME coupling inherent to a nanosystem is

$$\overline{g_{FME}^S}=\frac{1}{V}\int_V g_{FME}^S d^3r\sim\frac{1}{hc_{44}}\begin{pmatrix} \frac{-2r_er_m\overline{P_1L_1}Q_{44}^{(e)}Q_{44}^{(m)}}{(r_e+r_m)(r_e+\Lambda_e)(r_m+\Lambda_m)}-Q_{44}^{(e)}d_{15}^{(m)}\frac{r_e\overline{P_1L_1}}{(r_e+\Lambda_e)} \\ -Q_{44}^{(e)}d_{15}^{(m)}\frac{r_m\overline{P_1L_1}}{(r_m+\Lambda_m)}+Q_{44}^{(e)}\left(\widetilde{q}_{44}^{(m)}-2q_{44}^{(m)}\right)\frac{r_e\overline{P_1L_1L_3}}{(r_e+\Lambda_e)} \end{pmatrix} \tag{4.11}$$

It is seen from Eq.(4.11) that the coupling term has a pretty strong thickness dependence, namely its strength is inversely proportional to the film thickness *h*. It is seen that the influence of the flexoeffects decreases with either thickness increase or extrapolation length increase.

In order to generalize the result (4.9) to other symmetries as well as on both confined and bulk antiferromagnetics-ferroelectrics, we list the symmetry classes of antiferromagnetics-ferroelectrics, which have flexomagnetic effects (some $Q^{(m)}_{ijkl} \neq 0$), in the **Table 3**. Note, that all groups from the table can be the surface ones, i.e. they are listed in Eq.(1.4).

**Table 3.** Antiferromagnetics-ferroelectrics with flexomagnetic effect (multiferroics of the type I)

<table>
<tr><th rowspan="2">Point symmetry group</th><th rowspan="2">Magnetic symmetry group</th><th colspan="4">Number of the tensors nontrivial components *</th></tr>
<tr><th>flexo-magnetic</th><th>linear magneto-electric</th><th>piezo-magnetic</th><th>piezo-electric</th></tr>
<tr><td>mm2</td><td>mm2</td><td>13</td><td>2</td><td>3</td><td>5</td></tr>
<tr><td>4</td><td>4′</td><td>14</td><td>2</td><td>4</td><td>4</td></tr>
<tr><td rowspan="2">4mm</td><td>4mm</td><td>6</td><td>1</td><td>1</td><td rowspan="2">3</td></tr>
<tr><td>4′mm′</td><td>7</td><td>1</td><td>2</td></tr>
<tr><td>3m</td><td>3m</td><td>9</td><td>1</td><td>2</td><td>4</td></tr>
<tr><td>6</td><td>6′</td><td>6</td><td>0</td><td>2</td><td>4</td></tr>
<tr><td rowspan="2">6mm</td><td>6mm</td><td>5</td><td>1</td><td>1</td><td rowspan="2">3</td></tr>
<tr><td>6′mm′</td><td>3</td><td>0</td><td>1</td></tr>
</table>

Note, that not all antiferromagnetics-ferroelectrics are linear magnetoelectrics (e.g. 6′ and 6′mm′ symmetry has $f_{ij} \equiv 0$). So, the spontaneous FME coupling, induced either by the spatial confinement or by external bending, ***gives rise to the linear magnetoelectricity*** in the nanosized systems with e.g. 6′ and 6′mm′ symmetry.

## 5. Improper ferroelectricity induced by the flexo-magnetoelectric coupling in antiferromagnetics

It is rather important, that the FME effect (4.9) can induce the spontaneous polarization (i.e. **improper ferroelectricity**) in the antiferromagnetics. Equation of state for polarization vector follows from the variation of the free energy (2.1) $\delta F_V / \delta P_i = 0$. Variation of the Eq.(4.9) leads to the built-in field appearance in the right-hand side of the equation for polarization:

$$a^{(e)}_{11}(T)P_1 + a^{(e)}_{1jkl}P_jP_kP_l - E^{FME}_1 = 0, \tag{5.1a}$$

where the built-in field $E^{FME}_1$ induced by flexoeffects is given by expression:

$$E_1^{FME} = Q_{44}^{(e)} Q_{44}^{(m)} \frac{\partial^2 L_1}{\partial x_3^2} + Q_{44}^{(e)} d_{15}^{(m)} \frac{\partial L_1}{\partial x_3} - d_{15}^{(e)} Q_{44}^{(m)} \frac{\partial L_1}{\partial x_3} - 2\frac{Q_{44}^{(e)}}{c_{44}} \left(\tilde{q}_{44}^{(m)} - q_{44}^{(m)}\right) \frac{\partial (L_1 L_3)}{\partial x_3}. \quad (5.1b)$$

The built-in field (5.1b), proportional to the antimagnetization derivatives, can induce the polarization component $P_1$ in non-ferroelectric materials under the absence of external electric field, as shown in **Fig.5**. It is seen from the figure that nonzero polarization appears in the spatial regions, where the gradient of anti-magnetization vector **L** is pronounced.

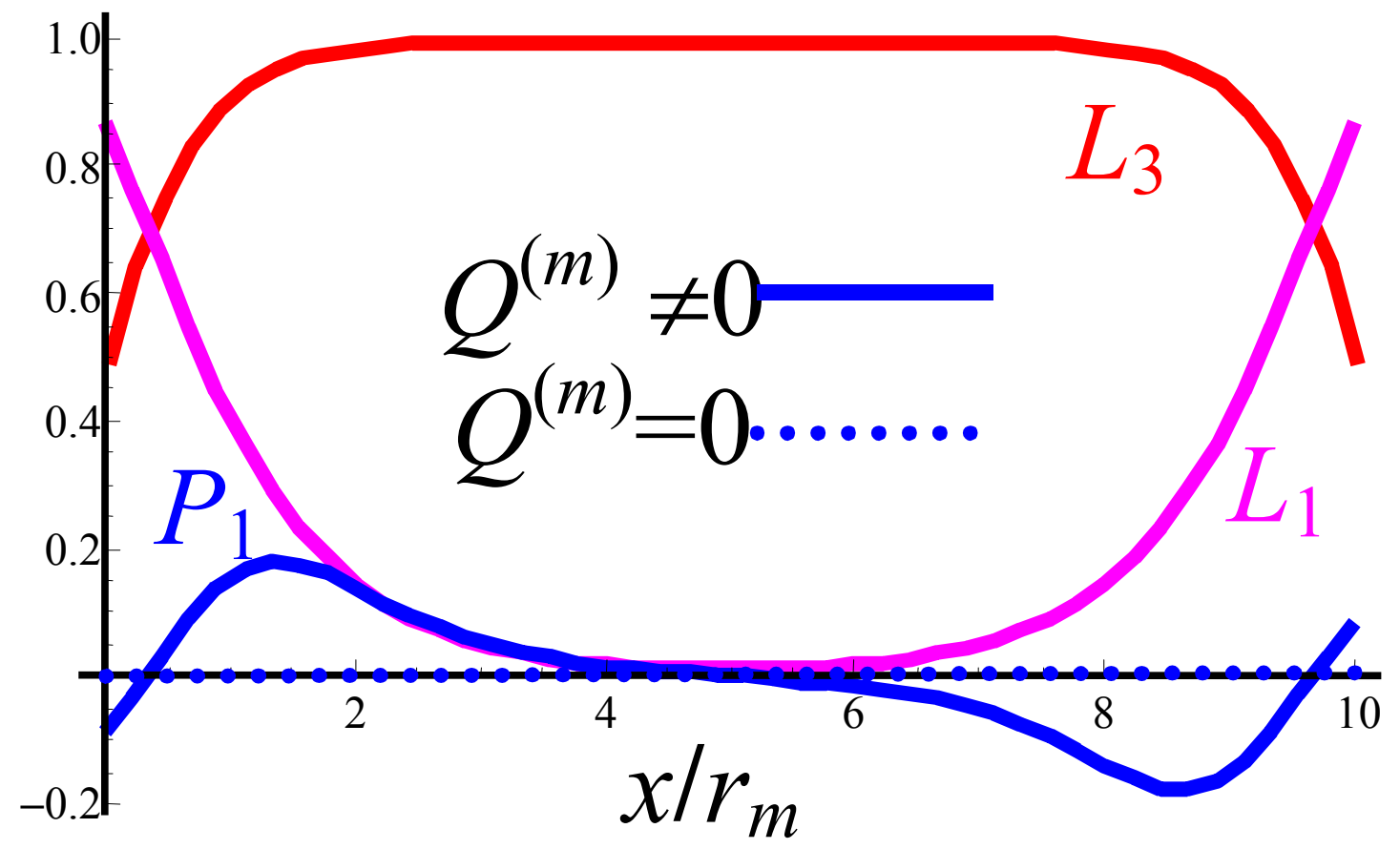

**Fig. 5.** Antimagnetization components $L_{1,3}(x)$ induce the improper polarization $P_1(x)$ in an antiferromagnetic. Dotted curve is zero polarization $P_1(x)=0$ calculated without flexomagnetic effect $Q^{(m)}=0$.

To be complete, we listed in the **Table 4** all the symmetry groups of antiferromagnetics, which can be improper ferroelectrics due to nonzero flexomagnetic effect (some of $Q_{ijkl}^{(m)} \neq 0$). Note, that seven of the groups have no linear magnetoelectric effect. So, the spontaneous FME coupling induced by the spatial confinement give rise to the ***size-dependent*** linear magnetoelectricity in the type II multiferroic nanosystems.

**Table 4.** Antiferromagnetics-improper ferroelectrics due to flexomagnetic effect (multiferroics of the type II)

| **Point symmetry group** | **Magnetic symmetry group** | **Number of the tensors nontrivial components *** | | | |
|---|---|---|---|---|---|
| | | flexo-magnetic | linear magneto-electric | piezo-magnetic | piezo-electric |
| $\bar{1}$ | $\bar{1}'$ | 54 | 9 | **0** | **0** |
| $2/m$ | $2/m'$ | 28 | 5 | **0** | **0** |
| | $2'/m$ | 26 | 4 | **0** | |
| 222 | 222 | 15 | 3 | 3 | 3 |
| *mmm* | *mmm'* | 13 | 2 | **0** | **0** |
| | *m'm'm'* | 15 | 3 | **0** | |
| $\bar{4}$ | $\bar{4}'$ | 14 | 2 | 4 | 4 |
| $\bar{4}2m$ | $\bar{4}2m$ | 7 | 1 | 1 | 2 |
| | $\bar{4}'2m'$ | 8 | 2 | 2 | |
| | $\bar{4}'2'm$ | 6 | 1 | 2 | |
| 422 | 422 | 8 | 2 | 1 | 1 |
| | 4’22’ | 7 | 1 | 2 | |
| 4/*m* | 4/*m’* | 14 | 2 | **0** | **0** |
| | 4’/*m’* | 14 | 2 | **0** | |
| 4/*mmm* | 4/*m’mm* | 6 | 1 | **0** | **0** |
| | 4/*m’m’m’* | 8 | 2 | **0** | |
| | 4’/*m’mm’* | 7 | 1 | **0** | |
| 32 | 32 | 11 | 2 | 2 | 2 |
| $\bar{3}$ | $\bar{3}'$ | 18 | 2 | **0** | **0** |
| $\bar{3}m$ | $\bar{3}'m'$ | 11 | 2 | **0** | **0** |
| | $\bar{3}'m$ | 9 | 1 | **0** | |
| $\bar{6}$ | $\bar{6}'$ | 12 | 2 | 2 | 2 |
| $\bar{6}m2$ | $\bar{6}m2$ | **3** | **0** | 1 | 1 |
| | $\bar{6}'m'2$ | 7 | 2 | 1 | |
| | $\bar{6}'m2'$ | 5 | 1 | 1 | |
| 622 | 622 | 7 | 2 | 1 | 1 |
| | 6’22’ | **3** | **0** | 1 | |
| 6/*m* | 6/*m’* | 12 | 2 | **0** | **0** |
| | 6’/*m* | **6** | **0** | **0** | |
| 6/*mmm* | 6/*m’mm* | 5 | 1 | **0** | **0** |
| | 6/*m’m’m’* | 7 | 2 | **0** | |
| | 6’/*mmm’* | **3** | **0** | **0** | |
| 23 | 23 | **5** | 1 | 1 | 1 |
| *m*3 | *m'*3 | **5** | 1 | **0** | **0** |
| $\bar{4}3m$ | $\bar{4}3m$ | **2** | **0** | **0** | 1 |
| | $\bar{4}'3m'$ | **3** | 1 | 1 | |
| 432 | 432 | **3** | 1 | **0** | **0** |
| | 4’32’ | **2** | **0** | 1 | |
| *m*3*m* | *m’*3*m’* | **3** | 1 | **0** | **0** |
| | *m’*3*m* | **2** | **0** | **0** | |

*) Note, that none of the groups from the table can be the surface one (see Eq. (1.4))

Using the proposed approach one can show that improper magnetization can be induced by the flexomagnetic coupling both in bulk and nanosized antiferroelectrics.

## 6. The influence of flexo-magnetoelectric effect on susceptibility

We suppose that magnetic field **H** and electric field **E** are applied and consider a thin antiferroelectric film of 4m′m′ symmetry and the same geometry as shown in **Fig.3a** (magnetic anisotropy axis along axis $x_3$).

Introducing ferromagnetic and antiferromagnetic order parameters as $\mathbf{M}=\left(\mathbf{M}^{(a)}+\mathbf{M}^{(b)}\right)/2$ $\mathbf{L}=\left(\mathbf{M}^{(a)}-\mathbf{M}^{(b)}\right)/2$ and using the solution of linear elastic problem one could exclude the nontrivial strain components by means of Legendre transformation and get the following expression for the renormalized free energy:

$$\tilde{F}_V=\int\limits_V\left(\tilde{g}_{FE}+\tilde{g}_{AFM}+\tilde{g}_{ME}+\tilde{g}_{FME}\right)d^3r\,, \tag{6.1a}$$

$$\tilde{F}_S=\int\limits_S d^2r\left(\frac{a_i^S}{2}P_i^2+\left(2K_S+\tilde{K}_S\right)(\mathbf{Mn})^2+\left(2K_S-\tilde{K}_S\right)(\mathbf{nL})^2\right). \tag{6.1b}$$

Here the ferroelectric contribution acquires the form:

$$\tilde{g}_{FE}=\frac{a_1^{(e)}(T)}{2}P_1^2+\left(\frac{a_{11}^{(e)}}{4}-\frac{\left(q_{12}^{(e)}\right)^2}{2c_{11}}\right)P_1^4+...+\left(\frac{g_{44}^{(e)}}{2}-\frac{\left(Q_{44}^{(e)}\right)^2}{2c_{44}}\right)\left(\frac{\partial P_1}{\partial x_3}\right)^2-P_1E_1 \tag{6.2}$$

The antiferromagnetic contribution is

$$\begin{aligned}\tilde{g}_{AFM}=&J\left(\mathbf{M}^2-\mathbf{L}^2\right)-2(\mathbf{H}\mathbf{M})+2K\left(M_3^2+L_3^2\right)+\tilde{K}\left(M_3^2-L_3^2\right)+\\&+2\left(\frac{g_{44}^{(m)}}{2}-\frac{\left(Q_{44}^{(m)}\right)^2}{2c_{44}}\right)\left(\left(\frac{\partial M_1}{\partial x_3}\right)^2+\left(\frac{\partial L_1}{\partial x_3}\right)^2\right)+2\left(\frac{g_{11}^{(m)}}{2}-\frac{\left(Q_{11}^{(m)}\right)^2}{2c_{11}}\right)\left(\left(\frac{\partial M_3}{\partial x_3}\right)^2+\left(\frac{\partial L_3}{\partial x_3}\right)^2\right)\\&+\left(\tilde{g}_{44}^{(m)}-\frac{\left(Q_{44}^{(m)}\right)^2}{c_{11}}\right)\left(\left(\frac{\partial M_1}{\partial x_3}\right)^2-\left(\frac{\partial L_1}{\partial x_3}\right)^2\right)+\left(\tilde{g}_{11}^{(m)}-\frac{\left(Q_{11}^{(m)}\right)^2}{c_{11}}\right)\left(\left(\frac{\partial M_3}{\partial x_3}\right)^2-\left(\frac{\partial L_3}{\partial x_3}\right)^2\right)\end{aligned} \tag{6.3}$$

The condition $J>0$ is necessary for antiferromagnetic state ($\mathbf{M}=0,\ \mathbf{L}\neq 0$) to be stable in zero magnetic field. The condition $J<0$ is necessary for ferromagnetic state ($\mathbf{L}=0,\quad \mathbf{M}\neq 0$) to be stable at arbitrary magnetic fields.

Note, that in Eq.(6.3) we neglected the bilinear coupling between components of magnetization, found by Dzyaloshinskii for some magnetic symmetries and responsible for a weak magnetism in antiferromagnetics [47].

Magnetoelectric energy is

$$\tilde{g}_{ME}=2f_{11}M_1P_1+2w_{111}M_1P_1^2-2\frac{q_{12}^{(e)}q_{11}^{(m)}}{c_{11}}P_1^2\left(M_3^2+L_3^2\right)-\frac{q_{12}^{(e)}\tilde{q}_{11}^{(m)}}{c_{11}}P_1^2\left(M_3^2-L_3^2\right). \tag{6.4}$$

Flexo-magnetoelectric coupling energy is:

$$\tilde{g}_{FME} = \begin{pmatrix} -2\frac{q_{12}^{(e)}Q_{11}^{(m)}}{c_{11}}P_1^2\frac{\partial M_3}{\partial x_3} - 2Q_{44}^{(m)}\frac{Q_{44}^{(e)}}{c_{44}}\frac{\partial P_1}{\partial x_3}\frac{\partial M_1}{\partial x_3} - \frac{Q_{44}^{(e)}}{c_{44}}d_{15}^{(m)}M_1\frac{\partial P_1}{\partial x_3} \\ -d_{15}^{(e)}\frac{Q_{44}^{(m)}}{c_{44}}P_1\frac{\partial M_1}{\partial x_3} - 2\frac{Q_{44}^{(e)}}{c_{44}}\left(\frac{\partial P_1}{\partial x_3}\right)\left(\left(q_{44}^{(m)} + \tilde{q}_{44}^{(m)}\right)M_1M_3 + \left(q_{44}^{(m)} - \tilde{q}_{44}^{(m)}\right)L_1L_3\right) \end{pmatrix} \qquad (6.5)$$

Note that the terms quadratic with respect to magnetization vector, $\sim M_iM_j\,\partial P_k/\partial x_l$ and $\sim L_iL_j\,\partial P_k/\partial x_l$, exist in all materials, since the flexoelectric tensor $Q_{ijkl}^{(e)}$ and magnetostriction tensor $q_{ijnp}^{(m)}$ have nonzero components for arbitrary symmetry. However, the terms linear with respect to magnetization vector, $\sim\frac{\partial P_1}{\partial x_3}\frac{\partial M_1}{\partial x_3}$ and $\sim P_1^2\frac{\partial M_3}{\partial x_3}$, appear at magnetic fields higher than the critical field of the spin-flop phase transition in the antiferromagnetic nanomaterial with nonzero flexomagnetic effect (some of $Q_{ijkl}^{(m)} \neq 0$), i.e. in for the symmetry groups listed in the **Tables 3,4**.

The free energy (6.1) could be used for the description of several different situations, namely:
(a) **L**≠0 and **M**=0 at magnetic field lower than the critical one; for this case only **P** and **L** are nonzero.
(b) **L**≠0 and **M**≠0 for magnetic field higher than the critical one, but lower than the spin-flop; for the case **P**, **L**, **M** are nonzero.
(c) **M**≠0 and **L**=0 for ferromagnetic phase at high magnetic field above the spin-flop phase transition; for the case **P** and **M** are nonzero.

To explore the magnetoelectric properties of the material, dielectric susceptibility and magnetoelectric tunability is most often measured experimentally. Average magnetization, polarization, linear dielectric susceptibility at different magnetic fields and magnetoelectric tunability can be calculated from the free energy (6.1).

The dependence of dielectric susceptibility and magnetoelectric tunability on the magnetic field is shown in **Fig.6.** It is seen that the effect of FME coupling between the polarization and magnetization on the tunability and dielectric susceptibility is very high. Namely in the absence of flexo-effects the tunability due to the quadratic ME coupling could not exceed one percent (see **Fig. 6b**), while the flexo-coupling leads to the tunability about 10-30% (see **Fig.6d**).

Note that previously [48] we considered only the quadratic ME effect contribution to the polarization, magnetization and susceptibilities as well as we neglected the flexoelectric effect.

The magnetoelectric tunability of $Er_2O_3$ nanoparticles with 2-10 nm size and amorphous-like structure was giant (up to 120%) at 275 K [49]. We expect that giant magnetoelectric tunability could be related with the size-dependent flexomagnetoelectricity.

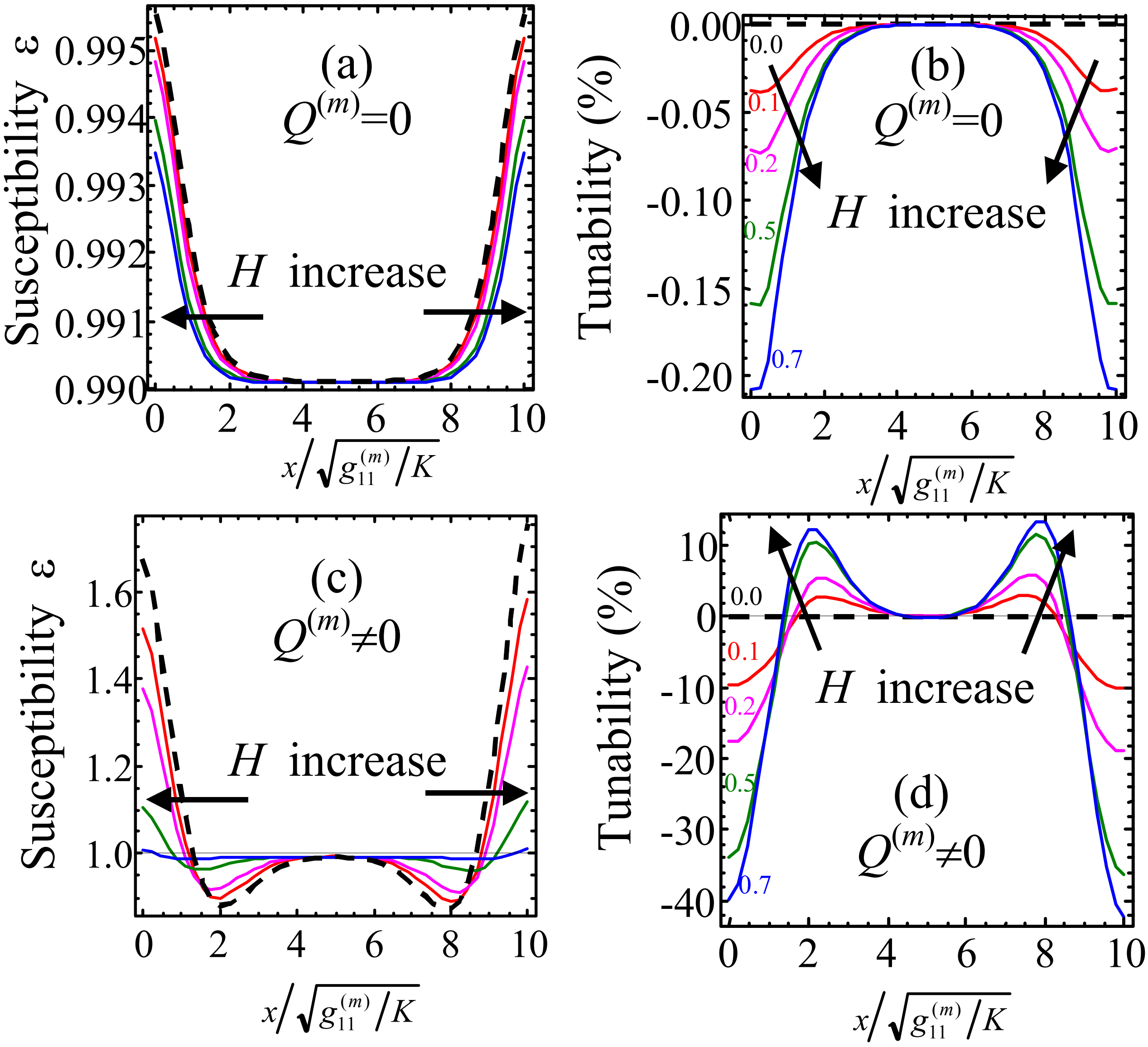

**Fig. 6.** Linear dielectric susceptibility (a,c) and magnetoelectric tunability (b,d) at increasing magnetic fields $H/(MK)$=0, 0.1, 0.2, 0.5, 0.7 (different curves) for cases of zero flexomagnetic coupling (a, b) and nonzero flexomagnetic coupling (c, d). Dashed curve is for zero magnetic field.

## 7. Discussion and summary

**To summarize**, we would like to underline the following:

1) Only 69 of the existing 90 bulk magnetic classes are flexomagnetic. Performed symmetry analyses showed that materials of 90 magnetic classes can be flexomagnetic, piezomagnetic and piezoelectric in the vicinity of the surface, due to the spontaneous symmetry lowering in the vicinity of surface. Here the local symmetry is changing the form of the flexomagnetic tensors. All possible surface magnetic classes have the number of 19 and can be obtained from the 90 bulk magnetic classes for the surface cuts 001, 010 and 100. Corresponding values of the flexoelectric and flexomagnetic tensors can be determined experimentally or calculated from first principles. Namely, typical values of flexoelectric tensor measured experimentally are $10^{-8}$ C/m for oxide dielectrics [10] and $10^{-6}$ C/m [7, 8] for ferroelectrics. In order to estimate the order of flexomagnetic coefficient we used the results of DFT calculations [17], where the flexomagnetic coefficient was calculated as about 1.95 $\mu_B$ Å for antiperovskite $Mn_3GaN$ as the result of a coupling between the strain gradient and magnetic dipole moment per Mn atom. Supposing that the unit cell of volume $6*10^{-29}$ $m^3$ bears three identical moments and using the values of fundamental constants ($\mu_B \approx 0.93*10^{-23}$ A $m^2$) we get the following value of the flexomagnetic coefficient ~$1.14*10^{-10}$ T m that couples the strain gradient ($m^{-1}$) and magnetic field (A/m). Using $Q^{(e)}$~$10^{-8}$ C/m, $Q^{(m)}$~$10^{-10}$ T m and size factors $r_{e,m}$~$10^{-9}$ m, $\Lambda_{e,m}$~$10^{-9}$ m and $s_u$~$10^{-11}$ $m^2$/N as the relevant elastic compliance, we obtain from Eq.(2.13) the estimation of the linear FME coupling coefficient $\frac{2r_e r_m s_u Q^{(m)} Q^{(e)}}{(r_e+r_m)(r_e+\Lambda_e)(r_m+\Lambda_m)} \sim 10^{-11}$ m/s, which is comparable with the linear ME coefficient value ~30 pm/s [50]. Some most promising multiferroics with a flexomagnetic effect are listed in the **Table 5**.

2) The flexomagnetic effect leads to the new type of spontaneous FME coupling in nanosized ferroelectrics-ferromagnetics and ferroelectrics-antiferromagnetics, where the gradients of polarization and magnetization vector are inevitably present due to the dominant role of the surface. The strength of the FME coupling is inversely proportional to the system characteristic size (e.g. nanotube radius or film thickness). Linear magnetoelectric effect can be induced by the linear FME effect in nanosized ferroelectrics-(anti)ferromagnetics, if one of its sizes is less than 5-10 nm. The FME coupling strongly affects the order parameters (polarization and magnetization vectors) spatial distributions as well the phase diagram of the nanosized multiferroics.

3) Improper ferroelectricity can be induced by the flexomagnetic-flexoelectric, flexomagnetic-piezoelectric, flexoelectric-piezomagnetic and/or piezomagnetic-piezoelectric couplings both in bulk and nanosized ferromagnetics and antiferromagnetics. In particular, pronounced maxima appear on the polarization distribution in the regions, where the gradients of the magnetization and/or anti-magnetization vectors exists (e.g. near the surfaces, where the vectors change their values due to the rotation). Similarly, improper magnetization can be induced by the flexomagnetic coupling both in bulk and nanosized ferroelectrics and antiferroelectrics. The main difference between nanosized and bulk materials is that improper polarization or magnetization are spontaneous for the nanosized case, while for a bulk ferromagnetic a pronounced magnetization (or polarization) vector gradient can be induced by external stresses as well as in the vicinity of domain walls.

4) The FME coupling induced by the spatial confinement give rise to the size-dependent linear magnetoelectricity in the considered multiferroic nanosystems. The FME coupling between the polarization and magnetization strongly influences the dielectric susceptibility and magnetoelectric tunability of nanosized multiferroics: in the absence of flexo-effects the tunability due to the quadratic ME coupling could not exceed one percent, while the flexo-coupling increases the tunability to 30%.

5) The flexomagnetic effect strongly increases the number of the magnetoelectric multiferroic materials of the type-I (ferroelectrics-(anti)ferromagnetics with two order parameters - spontaneous polarization and (anti)magnetization) and type-II (the materials with inhomogeneous spontaneous (anti)magnetization that induces the polarization, or vice versa). Therefore flexoeffects can be used for design of new multiferroics.

**Table 5.** Magnetic symmetry of some multiferroics with flexomagnetic effect

| **material** | **Magnetic symmetry** | **Multiferroic phase and type** | **Conditions of the multiferroic phase existence** | Ref. |
|---|---|---|---|---|
| $PbVO_3$ | 4mm | No | Above the Neel temperature | [51] |
| | 4′/m′mm′<br>(G type) | FE-AFM, type I | Below ~100–130 K<br>Thin films, | |
| $LiMPO_4$ with M = Fe, Co, Ni | mmm′ | AFM-im-FE-, type II | Below the Neel temperature (between 21.8 and 50 K), ferrotoroidic phase | [52] |
| $TbMn_2O_5$ | orthorhombic (space group Pbam) *mmm* | No | at room temperature | [53] |
| | m′m′2 | Yes | magnetic symmetry at low temperatures | |
| $BiFeO_3$ | *4′/m′mm′*<br>(G type) | FE-AFM, type I | Below the Neel temperature | [54] |
| $YMnO_3$, $InMnO_3$ | 6*mm*<br>($P6_3cm$) | AFM-im-FE, type II | noncollinear triangular antiferromagnetic spin configuration | [55] |
| $RMn_{1-x}Ga_xO_3$ (*R*=Ho,Y) | *6/mmm*<br>($P6_3/mmc$) | No | Above the Neel temperature | [56] |
| | 6*mm*<br>($P6_3cm$) | FE-AFM, type I | ferroelectric-(ferroelastic) phase | |
| | $P6'_3c'm$,<br>$P6'_3cm'$,<br>$P6_3cm$ | AFM-im-FE, type II | low temperature (antiferromagnetic) phases | |

*) FE-FM is ferroelectric-ferromagnetic, FE-AFM is ferroelectric-antiferromagnetic, AFM-im-FE antiferromagnetics improper ferroelectrics

1. Sh. M. Kogan, *Sov. Phys.—Solid State* **5,** 2069 (1964)

2. A.K. Tagantsev, *Phys. Rev B*, **34**, 5883 (1986)

3. A K. Tagantsev Electric polarization in crystals and response to thermal and elastic perturbations. *Phase Transitions.* **35** 119 (1991)

4 A.K. Tagantsev, Eric Courtens and Ludovic Arzel, Phys. Rev. B, **64**, 224107 (2001) (see Appendix)

5. W. Ma and L E Cross, *Appl. Phys. Lett.* **79,** 4420 (2001).

6. W. Ma and L E Cross, *Appl. Phys. Lett.* **81,** 3440 (2002).

7. W. Ma and L. E. Cross, Appl. Phys. Lett., **82**, 3293 (2003).